\documentclass{article}
\usepackage{arxiv}
\usepackage{graphicx}
\usepackage{mathptmx}
\usepackage[subrefformat=parens,labelformat=parens]{subcaption}
\usepackage{authblk}
\usepackage{amsmath}
\usepackage{amsfonts}
\usepackage{amssymb}
\usepackage{amsbsy}
\usepackage{amsthm}
\usepackage{comment}
\usepackage{amsmath}

\usepackage[ruled, vlined]{algorithm2e}
\usepackage{bbm}
\usepackage{wrapfig}
\usepackage{booktabs}
\usepackage{multirow}
\usepackage{wrapfig,lipsum,booktabs}
\usepackage{hyperref}

\begin{document}

\title{\Large 
Stochastic Molecular Reaction Queueing Network Modeling for 
In Vitro Transcription  
Process
}

\author[1]{Keqi Wang}
\author[  ~,1]{Wei Xie\thanks{Corresponding author: w.xie@northeastern.edu}}
\author[1]{Hua Zheng}
\affil[1]{Northeastern University, Boston, MA 02115}

\maketitle

\section*{ABSTRACT}
To facilitate a rapid response to pandemic threats, this paper focuses on developing a mechanistic simulation model for in vitro transcription (IVT) process, a crucial step in mRNA vaccine manufacturing. To enhance production and support industry 4.0, this model is proposed to improve the prediction and analysis of IVT enzymatic reaction network. It incorporates a novel stochastic molecular reaction queueing network with a regulatory kinetic model characterizing the effect of bioprocess state variables on reaction rates. The empirical study demonstrates that the proposed model has a promising performance under different production conditions and it could offer potential improvements in mRNA product quality and yield.

\section{INTRODUCTION}
\label{sec:intro}

In recent years, we have experienced several viral outbreaks, including the COVID-19 pandemic caused by the SARS-CoV-2 virus.
Vaccines are highly effective in stopping epidemics and pandemics, but traditional vaccine production methods are too slow to respond to new viral outbreaks \cite{gebre2021novel}. 
To overcome this issue, a rapid-response vaccine production platform is urgently needed.
Compared to traditional vaccines, mRNA vaccines offer several key benefits. First, they can be produced rapidly at a large scale, making mRNA vaccines ideal for responding to new viral outbreaks. Second, they are highly effective in preventing infection, as shown with the success of new mRNA vaccines developed for COVID-19. Third, mRNA vaccines are safe and well-tolerated, with few side effects reported.


The mRNA manufacturing involves key steps including RNA synthesis, purification, and formulation. Basically, the synthesis of RNA strands encodes the desired antigenic protein. For mRNA and self-amplifying RNA (saRNA) vaccine platforms, cell-free DNA-templated RNA synthesis is used for this purpose. It utilizes in vitro transcription (IVT) reactions that are often catalyzed by the T7 RNA polymerase enzyme (T7RNAP). The synthesized RNA substance is then purified and formulated with a delivery system (such as lipid nanoparticles), 
which are filled into vials or other containers to create the final vaccine product. 

This paper focuses on developing a mechanistic simulation model for the IVT process.  
Similar to the classical 
assembly lines, the IVT process synthesizes RNA chains based on the DNA templates. 
It has a complex reaction network 
involving many factors (such as enzymes, DNA templates, nucleoside triphosphates (NTPs), temperature, and pH) that interactively impact on the production outputs, including yield and product quality.
Developing a mechanistic simulation model for the IVT process is crucial in advancing the understanding of underlying mechanisms, identifying key reaction parameters, and optimizing the manufacturing process. 
The existing partial differential equation/ordinary differential equation (PDE/ODE) kinetic models for in vitro mRNA synthesis \cite{arnold2001kinetic,van2021quality} incorporate mechanisms from multiple phases, i.e., enzyme binding, initiation, elongation, and termination. They take into account the dynamic changes in enzyme activity, DNA template availability, and substrate concentration and also consider the effect of factors such as temperature, pH, and ionic strength.  However, the existing kinetics modeling approaches are typically deterministic and ignore the intrinsic stochasticity of the IVT process. 
While there exist stochastic simulation models for 
therapeutic protein production \cite{wang2019stochastic,xie2023stochastic} and mammalian cell culture \cite{zheng2023stochastic,wang2023metabolic}, a novel stochastic simulation model 
for the IVT process is currently lacking in the literature. 

To support mechanism learning and accelerate automation of mRNA vaccine manufacturing,
we propose \textit{a stochastic molecular reaction queueing network model} for the IVT process with the regulatory mechanism accounting for the impact of state variables (e.g., the concentrations of DNA template and enzyme, environmental conditions) on the reaction rates. We establish a relationship between outputs (e.g.,  RNA yield, and product critical quality attributes (CQAs)) with key input factors, such as initial NTP, {Magnesium (Mg) concentration, and buffer profile. 
The proposed model can improve the IVT process predictions through advancing the understanding of underlying reaction regulation mechanisms. 

Therefore, the objective of this study is to develop a stochastic 
mechanistic model to support IVT process prediction and decision making, which can improve product quality consistency and increase yield. The key contributions are threefold. First, we develop a molecular reaction queueing network model to simulate the RNA synthesis process, accounting for inherent stochasticity. Second, we create a regulatory mechanistic model characterizing the dynamic impact of bioprocess state on the reaction rates.
Third, the empirical study demonstrates that the proposed model 
can enhance mRNA product quality attributes and improve yield. 
Along with the advanced sensors, the proposed model has the potential to accelerate the development of flexible intensified mRNA vaccine manufacturing processes and support Industry 4.0.


The paper is organized as follows. We provide a brief introduction of the IVT process in Section~\ref{sec.IVTProcess}, and describe 
the state variables and their relationships in terms of mass balance and equilibrium Section~\ref{sec:IVT_State}. Then, we propose a stochastic molecular reaction queueing network and model the regulatory mechanism of reaction rates in Section~\ref{sec.IVT-StochasticModels}. This IVT process model is validated by using literature data and 
its performance is studied under various conditions in Section~\ref{sec:empirical}. Finally, we conclude the paper in Section~\ref{sec: conclusion}.


\section{IVT Process and Reaction Network}
\label{sec.IVTProcess}

This section reviews the mechanisms of the IVT process and the associated simplified reaction network.

\subsection{In Vitro Transcription Process}
\label{subsec:IVT}

RNA molecules can be produced through in vitro transcription, utilizing the RNA polymerase enzymes (e.g., T7 RNAP), DNA template, nucleoside triphosphates (NTPs) as substrates, and a transcription buffer solution containing Mg and other factors \cite{young1997modeling}.
The IVT process is typically divided into three stages: initiation, elongation, and termination \cite{cheetham1999structural,dousis2023engineered}; see an illustration in Figure~\ref{fig:enzymatic_networks}.
The DNA template includes the RNA polymerase (RNAP) promoter for transcription initiation, the code for the RNA molecule elongation, and the terminator sequence. 
RNAP is a group of enzymes that catalyzes the synthesis reactions of RNA molecules from a DNA template through the process of transcription 
\cite{lodish2008molecular}. T7 RNAP, derived from the T7 bacteriophage, is a popular enzyme used for IVT process in mRNA vaccine manufacturing \cite{tabor1985bacteriophage}. 

\begin{figure}[ht]
	\centering
	\includegraphics[width=0.8\textwidth]{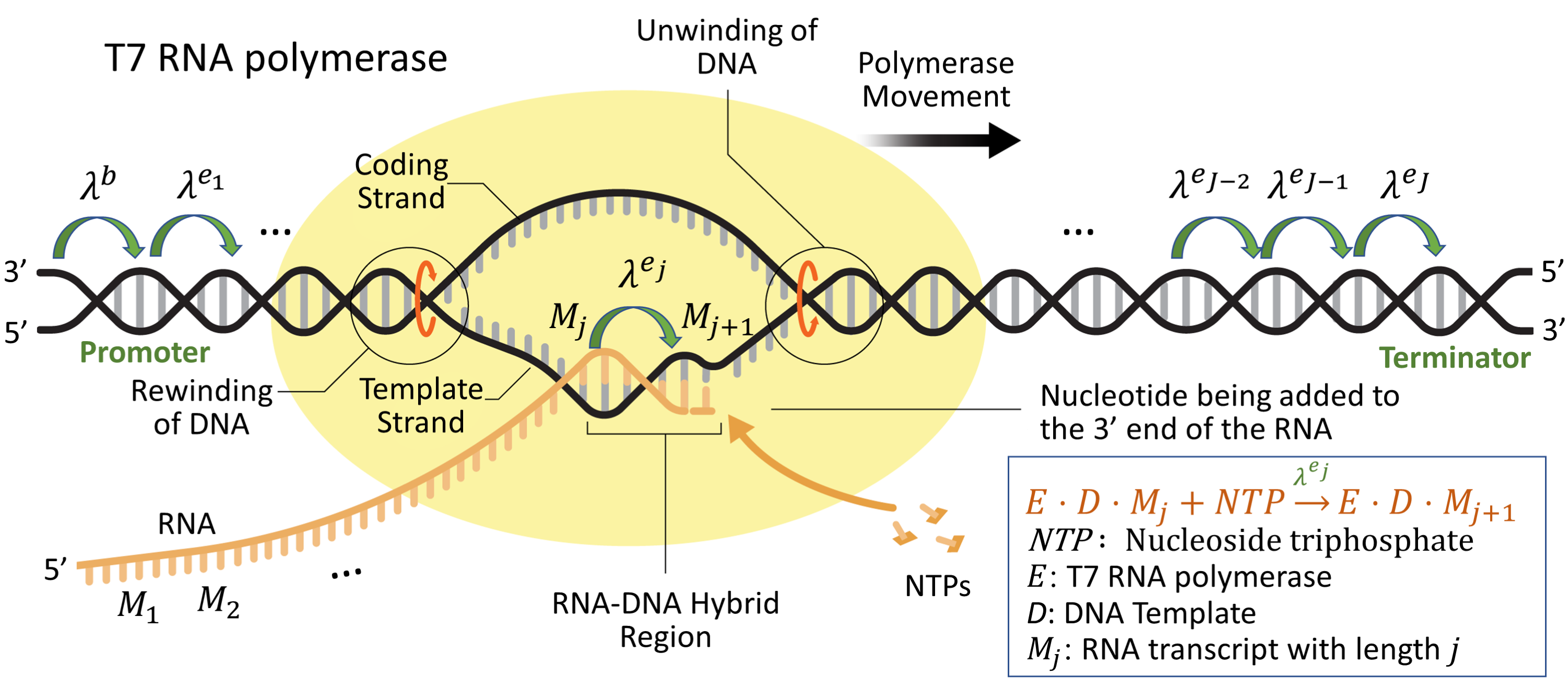}
 \vspace{-0.in}
	\caption{
	Topology of the mRNA synthesis mechanism with T7 RNAP in three stages of transcription (the plot is adapted from \protect\cite{dornell2021rna}). 
	}
	\label{fig:enzymatic_networks}
\end{figure}



\vspace{0.05in}

\noindent\textbf{(1) Binding and initiation:} 
Initiation is the first stage of the IVT process, during which T7 RNAP enzyme (denoted as $E$) binds to the promoter region of DNA template (denoted as $D$) to form the transcription initiation complex (IC) \cite{cheetham1999structural}. The promoter is a specific DNA sequence that signals RNAP where to bind upstream of a gene sequence. In the case of T7 RNAP, the initiator nucleotide is GTP, which is incorporated into the first position of the synthesized RNA molecule. Magnesium ions (Mg$^{2+}$) play a critical role in the IVT process by supporting the activity of RNAP, stabilizing the RNAP-DNA complex, and promoting the binding of NTPs, including ATP, CTP, GTP, and UTP, to the RNAP. The initiation of enzymatic RNA synthesis reaction network can be written as: 

\begin{equation}
    E + D + MgGTP \rightleftharpoons E\cdot D\cdot GTP + Mg {\rightarrow} E\cdot D\cdot M_1 +Mg 
             \nonumber
\end{equation}
where $M_1$ stands for an RNA transcript with length $1$ for simplification, $E \cdot D \cdot M_1$ represents an intermediate enzymatic complex containing the enzyme (E), DNA (D), and the nascent RNA $(M_1)$. 

\vspace{0.05in}

\noindent\textbf{(2) Elongation:} 
During the elongation stage, a stable and processive enzyme elongation complex (EC) is formed. RNA polymerase adds magnesium-complexed NTPs 
to the growing RNA chain based on the DNA coding sequence, following the rule of Watson-Crick base-pairing interactions. 
Each time a new NTP is added, one phosphodiester bond between the growing RNA molecule and the newly bound NTP is formed, leading to the release of one pyrophosphate ion (PPi) and one hydrogen ion (H) impacting on the pH level \cite{young1997modeling}. Thus, the overall enzymatic RNA synthesis reaction during the elongation stage of the IVT process can be described as:


\begin{equation}
    E\cdot D\cdot M_j + MgNTP {\rightleftharpoons} E\cdot D\cdot M_j\cdot NTP + Mg {\rightarrow} E\cdot D\cdot M_{j+1} + PPi + H + Mg
    \nonumber
\end{equation}
where $M_j$ stands for an RNA transcript with length $j$ and $PPi$ represents inorganic pyrophosphate. 
\vspace{0.05in}

\noindent\textbf{(3) Termination:} 
At the termination stage of transcription, the interaction between RNAP and DNA template is disrupted upon encountering a terminator sequence or signal. This halts the addition of complementary NTPs to the RNA strand, and the RNA transcript is released, marking the end of In-Vitro transcription process.
The enzymatic RNA synthesis reaction during the termination stage can be expressed as:

\begin{equation}
    E\cdot D\cdot M_J {\rightarrow} E+D+M_J
    \nonumber
\end{equation}
where $M_J$ stands for the full-length RNA transcript with length $J$. 

The IC is unstable, which can produce short RNA transcripts known as abortive transcripts with 2 to 10 nucleotides in length, in a process called \textit{abortive cycling} \cite{dousis2023engineered}. In addition, RNA degradation is a common issue that can have a negative impact on the yield and the RNA product CQAs. 
This process involves the breakdown of RNA molecules into smaller components, 
which can be caused by various factors, including inappropriate pH level, high temperature, buffer conditions, 
and contamination.


\subsection{Critical Process Parameters}
\label{subsec:CPPs}

The RNA product quality is associated with the percentage of impurities coming from the synthesis of short abortive RNAs and the degradation of the full length RNAs. 
Based on the study in \cite{van2021quality}, Table~\ref{table:CPP_and_CQA} summarizes the critical process parameters (CPPs) that impact the quality and yield of IVT production process. 
These CPPs include the pH level in the reactor, as well as the concentrations of magnesium, DNA template, T7 RNAP, and NTPs. 
They are actively involved in the regulatory reaction network of the IVT process (see Figure~\ref{fig:regulatory_networks}). The regulatory mechanisms discussed below will be integrated into the proposed stochastic reaction network model in Section~\ref{sec.IVT-StochasticModels}.

\begin{table}[h!]
\caption{The criticality of CPPs on the product CQAs. The magnitude of the impact is rated from 0 (low) to 3 (high). The direction and type of CPP-CQA relationship is characterized either by a positive impact labeled with plus $``+"$, a negative impact labeled with a minus $``-"$, or a peak behavior whereby the CQA increases with increasing the CPP reaches a peak and then decreases, labeled with plus-minus $``\pm"$. }

\label{table:CPP_and_CQA}
\centering
\begin{tabular}{|l|l|l|}
\hline
Process parameter             & RNA integrity and impurities & RNA yield \\ \hline
pH in transcription reactor   & $\pm 2$                             & $\pm 3$   \\ \hline
Total Mg concentration        & $\pm 2$                                  & $\pm 3$   \\ \hline
DNA template concentration    & 0                                         & $+2$      \\ \hline
T7 RNAP concentration         & $\pm 1$                             & $+3$      \\ \hline
Total NTP concentration       & $+2$                                & $\pm 3$   \\ 
\hline
\end{tabular}
\end{table}

\vspace{-0.in}

\begin{itemize}
    \item [(1)] \textbf{pH:} 
    The pH, reflecting the concentration of hydrogen (H) ions, plays a critical role in enzyme binding to DNA template, enzyme activity, and RNA degradation rate during the IVT process. T7 RNAP exhibits optimal activity at a pH range of 7.9 $\sim$ 8.1 \cite{kartje2021revisiting}, although it can still function within a pH range of 7.3 $\sim$ 8.3. Deviations from this range may potentially decrease enzyme activity, negatively affecting the yield and RNA product quality. Therefore, it is crucial to continuously monitor and control the pH level during the IVT process and evaluate its impact.
    \item [(2)] \textbf{Magnesium:} Mg is an essential cofactor in the IVT reactions. It plays a critical role in stabilizing T7RNAP-DNA complex and promoting RNA synthesis \cite{vernon1988role}. However, a too high concentration of $\mbox{Mg}^{2+}$ also favors RNA degradation.
    \item [(3)] \textbf{PPi:} 
    As PPi is continuously produced during the IVT process, it can result in the formation and precipitation of magnesium pyrophosphate (Mg$_2$PPi). This can reduce the available free $\mbox{Mg}^{2+}$ ions in the solution and ultimately lead to a decrease in the IVT process yield. To avoid this, an enzyme called inorganic pyrophosphatase (iPPase) \cite{tersteeg2022purification} can be utilized to hydrolyze the PPi, preventing the formation of magnesium pyrophosphate and ensuring that magnesium ions stay in the solution, thereby optimizing T7 RNAP enzyme activity.

    \item [(4)] \textbf{T7 RNAP, DNA template, and NTPs:} The concentrations of these elements are critical as they serve as key raw materials to support the RNA synthesis process. Insufficient concentrations and inappropriate proportions of these key components can result in low yield and poor quality of RNA product. Therefore, it is essential to carefully control and maintain the concentrations of DNA template, T7 RNAP, and NTPs during the IVT process.
\end{itemize}

To ensure consistent RNA product quality attributes and improved yield, monitoring and regulating the CPPs in the IVT process is essential. For instance, the pH level, influenced by hydrogen generation, affects enzyme binding to DNA template, enzyme activity, and RNA degradation rate; see the reaction network illustration in Figure~\ref{fig:regulatory_networks}. Similarly, while magnesium plays a critical role in promoting RNA synthesis, the generation of PPi can reduce the free $\mbox{Mg}^{2+}$ concentration and inhibit RNA synthesis.
The control strategies, such as mechanistic hybrid model-based reinforcement learning \cite{zheng2021policy}, can support \textit{end-to-end IVT process control} so that these CPPs are optimized accounting for complex interactions and long-term effects.

\begin{figure}[ht]
	\centering
	\includegraphics[width=0.65\textwidth]{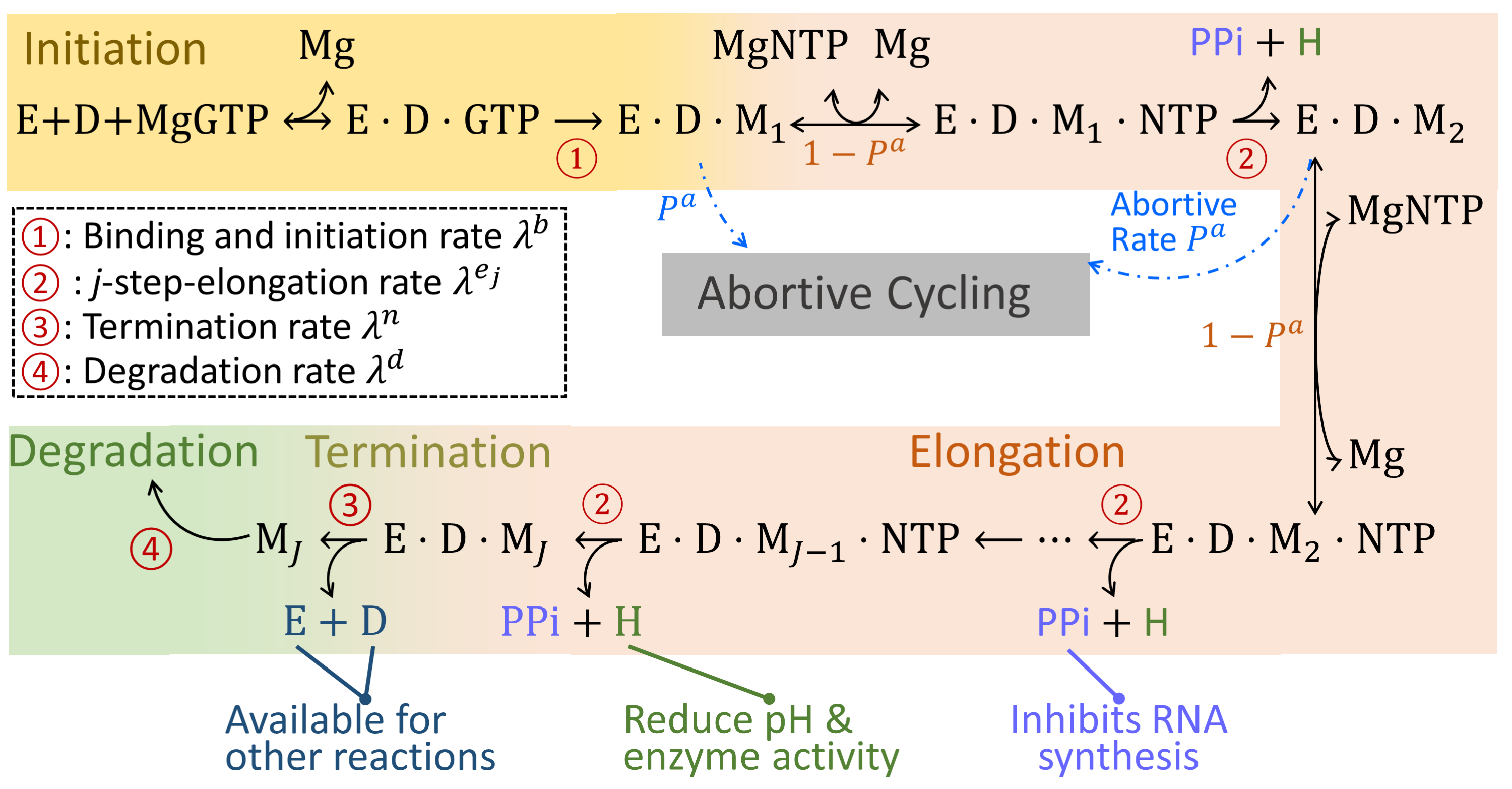}
 \vspace{-0.in}
	\caption{
	A simple illustration of the regulatory reaction network for the IVT process. 
	}	\label{fig:regulatory_networks}
\end{figure}

\section{State Variables}
\label{sec:IVT_State}
\begin{sloppypar}

In this section, we identify the IVT production system state variables of interest and present the mass balance and equilibrium equations as the constraints to the concentrations of different media components. 
Based on the information from Table~\ref{table:CPP_and_CQA} and {
\cite{van2021quality}}, at any time $t$, the system \textit{state variables} include the concentrations of molecules: T7 RNAP enzyme $[\mbox{T7RNAP}]_t$, DNA template $[\mbox{DNA}]_t$, free species $\pmb{s}^{free}_t$, complexes  $\pmb{s}^{comp}_t$, total $\pmb{s}^{tot}_t$, synthesized full-length mRNA $[\mbox{RNA}]_t$ and undesirable impurities $[\mbox{Impurity}]_t$:
\begin{equation*}
    \pmb{s}_t = \left\{[\mbox{T7RNAP}]_t, [\mbox{DNA}]_t, {\pmb{s}^{free}_t}^\top, {\pmb{s}^{comp}_t}^\top, {\pmb{s}^{tot}_t}^\top, [\mbox{RNA}]_t, [\mbox{Impurity}]_t \right\}^\top,
\end{equation*}
where the bracket $[\cdot]$ represents the concentration.

The main free species presented in the solution affecting transcription and degradation kinetics include: Mg$^{2+}$, NTP$^{4-}$ (i.e., ATP$^{4-}$, UTP$^{4-}$, CTP$^{4-}$, and GTP$^{4-}$), H$^+$, HEPES$^-$ (buffer) and PPi$^{4-}$, i.e., 
\begin{equation*}
    \pmb{s}_t^{free} = \left\{[\mbox{Mg}]_t, [\mbox{ATP}]_t, [\mbox{UTP}]_t, [\mbox{CTP}]_t, [\mbox{GTP}]_t, [\mbox{H}], [\mbox{HEPES}]_t, [\mbox{PPi}]_t \right\}^\top.
\end{equation*}
 These 8 free solution components can form the 22 complexes:  HATP$^{3-}$, HUTP$^{3-}$, HCTP$^{3-}$, HGTP$^{3-}$, MgATP$^{2-}$, MgUTP$^{2-}$, MgCTP$^{2-}$, MgGTP$^{2-}$, Mg$_2$ATP, Mg$_2$UTP, Mg$_2$CTP, Mg$_2$GTP, MgHATP$^-$, MgHUTP$^-$, MgHCTP$^-$, MgHGTP$^-$, MgPPi$^{2+}$, Mg$_2$PPi, HPPi$^{3-}$, H$_2$PPi$^{2-}$, MgHPPi$^-$ and HHEPES. At any time $t$, we represent the set of complex's concentrations as a vector $\pmb{s}_t^{comp}$. 
Then the total concentration of each species, defined as the sum of its free ion concentrations and the concentrations of its complexes, becomes:
\begin{equation*}
    \pmb{s}_t^{tot} = \left\{ [\mbox{Mg}]_t^{tot}, [\mbox{ATP}]_t^{tot}, [\mbox{UTP}]_t^{tot}, [\mbox{CTP}]_t^{tot}, [\mbox{GTP}]_t^{tot}, [\mbox{H}]_t^{tot}, [\mbox{PPi}]_t^{tot}, [\mbox{HEPES}]_t^{tot}\right\}^\top.
\end{equation*}
\end{sloppypar}

\begin{sloppypar}

In any short time period, we assume that the system is in equilibrium. That means the mass balance (outlined in Table~\ref{tab:MassBalance}) and equilibrium equations (outlined in Table~\ref{tab:Equilibrium}) are used to describe the concentration constrains on the chemical species.
Basically, the mass balance equations given in Table~\ref{tab:MassBalance} describe the conservation of mass for each species presented in the system. Each equation defines the relationship between the total concentration of a species in the system, its free ion concentration, and the concentrations of its complexes. For example, in Equation~\textbf{M1}, the total concentration of magnesium ions is equal to the sum of the concentration of free magnesium ions, magnesium ions bound to ATP, UTP, CTP, GTP, and their corresponding hydrolyzed products, and the concentrations of various magnesium-containing complexes.

\begin{table}[h!]
\centering
{
\caption{The mass balance equations (adapted from \protect\cite{van2021quality})}
\label{tab:MassBalance}
\begin{tabular}{|l|l|}
\hline
\multirow{3}{*}{\textbf{M1}} & \multirow{3}{*}{\begin{tabular}[c]{@{}l@{}}$[\mbox{Mg}]^{tot} = [\mbox{Mg}] + [\mbox{MgATP}] + [\mbox{MgUTP}] + [\mbox{MgCTP}] + [\mbox{MgGTP}] + 2 \times [\mbox{Mg$_2$ATP}]$\\ 
                        $ + 2 \times [\mbox{Mg$_2$UTP}]  + 2 \times [\mbox{Mg$_2$CTP}] + 2 \times [\mbox{Mg$_2$GTP}]  + [\mbox{MgHATP}] + [\mbox{MgHUTP}]$\\ 
                        $+ [\mbox{MgHCTP}] + [\mbox{MgHGTP}]  + [\mbox{MgPPi}] + 2 \times [\mbox{Mg$_2$PPi}] + [\mbox{MgHPPi}]$
                        \end{tabular}} \\
                    &   \\
                    &   \\ \hline
\textbf{M2} & $[\mbox{ATP}]^{tot} = [\mbox{ATP}] + [\mbox{MgATP}] + [\mbox{Mg$_2$ATP}] + [\mbox{MgHATP}] + [\mbox{HATP}]$  \\ \hline
\textbf{M3} & $[\mbox{UTP}]^{tot} = [\mbox{UTP}] + [\mbox{MgUTP}] + [\mbox{Mg$_2$UTP}] + [\mbox{MgHUTP}] + [\mbox{HUTP}]$  \\ \hline                                                             
\textbf{M4} & $[\mbox{CTP}]^{tot} = [\mbox{CTP}] + [\mbox{MgCTP}] + [\mbox{Mg$_2$CTP}] + [\mbox{MgHCTP}] + [\mbox{HCTP}]$  \\ \hline   
\textbf{M5} & $[\mbox{GTP}]^{tot} = [\mbox{GTP}] + [\mbox{MgGTP}] + [\mbox{Mg$_2$GTP}] + [\mbox{MgHGTP}] + [\mbox{HGTP}]$  \\ \hline   \multirow{2}{*}{\textbf{M6}} & \multirow{2}{*}{\begin{tabular}[c]{@{}l@{}} $[\mbox{H}]^{tot} = [\mbox{H}] + [\mbox{MgHATP}] + [\mbox{MgHUTP}] + [\mbox{MgHCTP}] + [\mbox{MgHGTP}] + [\mbox{HATP}] + [\mbox{HUTP}]$\\ 
                        $+ [\mbox{HCTP}] + [\mbox{HGTP}] + [\mbox{HPPi}] + 2 \times [\mbox{H$_2$PPi}] + [\mbox{MgHPPi}] + [\mbox{HHEPES}]$
                        \end{tabular}} \\
                    &   \\ \hline
\textbf{M7} & $[\mbox{PPi}]^{tot} = [\mbox{PPi}] + [\mbox{MgPPi}] + [\mbox{Mg$_2$PPi}] + [\mbox{HPPi}] + [\mbox{H$_2$PPi}] + [\mbox{MgHPPi}]$  \\ \hline
\textbf{M8} & $[\mbox{HEPES}]^{tot} = [\mbox{HEPES}] + [\mbox{HHEPES}]$  \\ \hline
\end{tabular}
}
\end{table}

The equilibrium equations given in Table~\ref{tab:Equilibrium} represent the chemical equilibrium conditions for the reactions involving various species in the system. Each equation describes an equilibrium constant for the reaction involving the species on the left- and right-hand sides of the equation. For example, in Equation~\textbf{E1}, 
the equilibrium constant $K_{eq,1}$ describes the equilibrium between ATP and its protonated form, HATP, in the presence of hydrogen ions.

\begin{table}[h!]
\centering
{
\caption{The equilibrium equations (adapted from \protect\cite{van2021quality})}
\label{tab:Equilibrium}
\begin{tabular}{|l|l||l|l|}
\hline
\textbf{E1}  &  $[\mbox{H}][\mbox{ATP}] = K_{eq,1}[\mbox{HATP}]$   & \textbf{E2}  & $[\mbox{H}][\mbox{UTP}] = K_{eq,2}[\mbox{HUTP}]$    \\ \hline
\textbf{E3}  & $[\mbox{H}][\mbox{CTP}] = K_{eq,3}[\mbox{HCTP}]$    & \textbf{E4}  & $[\mbox{H}][\mbox{GTP}] = K_{eq,4}[\mbox{HGTP}]$    \\ \hline
\textbf{E5}  & $[\mbox{Mg}][\mbox{ATP}] = K_{eq,5}[\mbox{MgATP}]$  & \textbf{E6}  & $ [\mbox{Mg}][\mbox{UTP}] = K_{eq,6}[\mbox{MgUTP}]$   \\ \hline
\textbf{E7}  & $[\mbox{Mg}][\mbox{CTP}] = K_{eq,7}[\mbox{MgCTP}]$  & \textbf{E8}  & $[\mbox{Mg}][\mbox{GTP}] =  K_{eq,8}[\mbox{MgGTP}]$   \\ \hline
\textbf{E9}  & $[\mbox{Mg}][\mbox{MgATP}] = K_{eq,9}[\mbox{Mg$_2$ATP}]$ & \textbf{E10}  & $[\mbox{Mg}][\mbox{MgUTP}] = K_{eq,10}[\mbox{Mg$_2$UTP}]$  \\ \hline
\textbf{E11} & $[\mbox{Mg}][\mbox{MgCTP}] = K_{eq,11}[\mbox{Mg$_2$CTP}]$ & \textbf{E12} & $[\mbox{Mg}][\mbox{MgGTP}] = K_{eq,12}[\mbox{Mg$_2$GTP}]$ \\ \hline
\textbf{E13} & $[\mbox{Mg}][\mbox{HATP}] = K_{eq,13}[\mbox{MgHATP}]$  & \textbf{E14} & $[\mbox{Mg}][\mbox{HUTP}] =  K_{eq,14}[\mbox{MgHUTP}] $\\ \hline
\textbf{E15} & $[\mbox{Mg}][\mbox{HCTP}] = K_{eq,15}[\mbox{MgHCTP}]$  & \textbf{E16} & $[\mbox{Mg}][\mbox{HGTP}] = K_{eq,16}[\mbox{MgHGTP}]$ \\ \hline
\textbf{E17} & $[\mbox{Mg}][\mbox{PPi}] = K_{eq,17}[\mbox{MgPPi}]$   & \textbf{E18} & $[\mbox{Mg}][\mbox{MgPPi}] = K_{eq,18}[\mbox{Mg$_2$PPi}]$  \\ \hline
\textbf{E19} & $[\mbox{H}][\mbox{PPi}] = K_{eq,19}[\mbox{HPPi}]$   & \textbf{E20} & $[\mbox{H}][\mbox{HPPi}] = K_{eq,20}[\mbox{H$_2$PPi}]$   \\ \hline
\textbf{E21} & $[\mbox{Mg}][\mbox{HPPi}] = K_{eq,21}[\mbox{MgHPPi}]$  & \textbf{E22} &  $[\mbox{H}][\mbox{HEPES}] = K_{eq,22}[\mbox{HHEPES}]$ \\ \hline
\end{tabular}
}
\end{table}

By solving these mass balance equations 
and equilibrium equations, 
the concentrations of any group and equilibrium constants can be used to calculate the concentrations of the other two groups. Typically, the initial concentrations of the total group are known except for $[\mbox{H}]_t^{tot}$. However, the concentration of free hydrogen ions $[\mbox{H}]_t$ can be determined by using the measured pH value, i.e., $\mbox{pH} = -\log([\mbox{H}])$. It should be noted that the choice of free components and complexes is dependent on the specific IVT reaction and can be adapted based on the particular experimental setup. The modeling approach presented in this paper can be extended to incorporate additional or alternative components.
\end{sloppypar}

\vspace{0.1in}

\section{Stochastic Molecular Reaction Queueing Network}
\label{sec.IVT-StochasticModels}


Building upon recent research studies  \cite{clement2020stochastic,kloska2022queueing}, 
this section aims to develop a stochastic molecular reaction network model for the enzymatic IVT process using queueing theory. 
In Section~\ref{subsec:Queue}, we model the IVT reaction process 
illustrated in Figure~\ref{fig:regulatory_networks}
as a queueing network 
producing the RNA molecule product. 
The queueing network includes multiple stages 
with the reaction rates (i.e., the number of molecular reactions occurring in unit time) depending on the state $\pmb{s}_t$.
Then, we model the state transition for $\pmb{s}_t$ characterizing the bioprocess macro-kinetics in Section~\ref{subsec:stateTransition}.



\subsection{Reaction Rate Modeling for Stochastic Molecular Reaction Queueing Network}
\label{subsec:Queue}

The IVT process queueing network is composed of multiple steps: binding and initiation, abortive cycling, elongation, termination, and degradation. The proposed reaction regulation mechanistic model  leverages the information from previously published studies \cite{akama2012multiphysics,arnold2001kinetic,van2021quality}. 
After the binding and initiation stage, the initiated enzymatic complexes have two possible routes to follow: abortive cycling or elongation; see Figure~\ref{fig:regulatory_networks}. In abortive cycling, the initiated enzymatic complexes prematurely release abortive RNA transcripts, before proceeding to the elongation stage. 
The probability of having an initiated complex entering abortive cycling, denoted by $P^a(\pmb{s}_t)$, 
is influenced by the state variables, such as the pH in the transcription reactor and the concentrations of NTPs. 
The abortive RNA transcripts are considered as impurity that reduces IVT product quality as well as yield. If the initiated enzymatic complex successfully proceeds beyond the 
initiation stage, 
it enters the elongation stage. During this stage, the T7 RNAP enzyme moves along the DNA template, synthesizing the RNA transcript. 


\begin{itemize}

\item[(1)] \textbf{Binding and initiation:} 
The synthesized RNA with length $j$, denoted as $M_j$, is treated as entities in the queueing system, 
where $j = 1, 2,\dots, J$. 
The generation rate of the nascent RNA $M_1$ depends on the DNA template and T7 RNAP enzyme binding rate, denoted by $\lambda^b(\pmb{s}_t)$. That means the DNA template and RNA polymerase are occupied, and 
GPT is consumed as the entities arriving at a rate of $\lambda^b(\pmb{s}_t)$. 
The reaction rate of the binding and initiation phase $\lambda^b(\pmb{s}_t)$ is modeled as 
    \begin{equation}
    \label{eq:binding_rate}
        \lambda^b(\pmb{s}_t) = V_{max,b} \times \frac{[\mbox{DNA}]_t}{K_{M,DNA} + [\mbox{DNA}]_t} \times \frac{[\mbox{MgGTP}]_t}{K^b_{M,MgGTP} + [\mbox{MgGTP}]_t} 
        \times \prod_{i \in \{\text{MgATP, MgUTP, MgCTP}\}} \frac{K_{I,i}}{K_{I,i} + [i]_t},   
    \end{equation}
    where the parameters $K_{I,\cdot}$, $K_{M,\cdot}$, and $V_{max,\cdot}$ represent the Michaelis-Menten (MM) inhibition constant, the affinity constant, and the maximum specific reaction rate respectively.     
    
    The binding and initiation rate in Equation~(\ref{eq:binding_rate}) is formulated as the product of three terms: a) the rate of promoter DNA binding, which is influenced by the concentration of DNA; b) the rate of initial GTP-Mg complex binding, which is influenced by the concentration of GTP-Mg complex; and c) a term that takes into account the competition among substrate NTP-Mg complexes. The latter term is given by the product of inverse MM constants ($K_{I,i}$) and the concentration of the competing NTP-Mg complexes, denoted by $[i]_t$, where $i$ belongs to the set of substrate NTP-Mg complexes including MgATP, MgUTP, and MgCTP. By appropriately quantifying the impact of each factor on the binding and initiation rate, Equation~\eqref{eq:binding_rate} can be used to predict the effect of changes in the concentrations of the various species on the efficiency of RNA synthesis.
    



\item[(2)] 
\begin{sloppypar} 
\textbf{Abortive Cycling:} The production of short RNA or abortive transcripts can reduce RNA yield and purity \cite{dousis2023engineered}. To account for this, we incorporate the routing probability of abortive cycling into the model and denote it as $P^a(\pmb{s}_t)$. 
The synthesized abortive RNA transcripts are considered impurities and reduce RNA product integrity.
As shown in Table~\ref{table:CPP_and_CQA}, the pH level in the transcription reactor and the concentrations of NTPs are the CPPs affecting the abortive RNA and impurity generation. 
Since the closed-form expression for the routing probability of abortive cycling, commonly referred to as the abortive rate, is not available in the existing literature, inspired from the work in \cite{xie2022interpretable}, we constructed a regulation mechanstic model to characterize the impact of pH (i.e., [H] and [OH]) and the concentrations of NTP-Mg complexes (i.e., [MgATP], [MgUTP], [MgCTP], [MgGTP]) on the abortive cycling routing probability $P^a(\pmb{s}_t)$, i.e.,
\begin{align}
    P^a(\pmb{s}_t) = f([\mbox{H}]_t,[\mbox{OH}]_t,[\mbox{MgATP}]_t,[\mbox{MgUTP}]_t,[\mbox{MgCTP}]_t,[\mbox{MgGTP}]_t) 
     = \frac{1}{1+e^{-(\kappa_t-1)}},
    \nonumber
\end{align}
where $\kappa_t = \phi_1 [\mbox{H}]_t + \phi_2 [\mbox{OH}]_t + \phi_3 [\mbox{MgATP}]_t + \phi_3 [\mbox{MgATP}]_t + \phi_4 [\mbox{MgUTP}]_t + \phi_5 [\mbox{MgCTP}]_t + \phi_6 [\mbox{MgGTP}]_t$ and the parameters $\phi_1$, $\phi_2$, $\phi_3$, $\phi_4$, $\phi_5$, $\phi_6$ characterize the influence from each CPP input.
We assume no delay in releasing the occupied DNA template and enzyme during abortive cycling, 
i.e., making resources immediately available for other reactions. 
\end{sloppypar}
\item[(3)] \textbf{Elongation:}
In the elongation stage, the T7 RNAP enzyme moves along the DNA template, synthesizing the RNA transcript. At each elongation step, 
one NTP unit is added into the growing transcript, i.e., the nascent RNA chain with length $j$ having an
increment rate denoted by $\lambda^{e_j}(\pmb{s}_t)$. This rate depends on the state variables $\pmb{s}_t$, such as reactant concentrations and environmental conditions. As the T7 RNAP enzyme progresses, the added NTP forms complementary base pairs with the DNA template, effectively elongating the RNA chain. Throughout this stage, both the DNA template and enzyme remain occupied, ensuring that the elongation process continues uninterrupted until the RNA transcript is complete.
The per-step-elongation rate depends on the concentration of the correct NTP-Mg complex, denoted by $[i]_t$ 
with $i \in \{\text{MgATP, MgUTP, MgCTP, MgGTP}\}$, as well as the concentrations of incorrect NTP-Mg complexes with $q \neq i$.
We model the elongation rate as,
    \begin{equation}
    \label{eq:elongation_rate}
        \lambda^{e_j}(\pmb{s}_t) = V_{max,i} \times \frac{[i]_t}{K_{M,i} + [i]_t}  
        \times \prod_{q \in \{\text{MgATP, MgUTP, MgCTP, MgGTP}\} \& q \neq i} \frac{K_{I,q}}{K_{I,q} + [q]_t}.   
    \end{equation}

    Thus, the elongation rate in Equation~(\ref{eq:elongation_rate}) is modeled as the product of two terms: a) the rate of type $i$-th NTP-Mg complex associated with the ternary complex of T7 RNAP, DNA, and RNA, which is influenced by the concentration of $[i]_t$; and b) a term that takes into account the competition among other substrate NTP-Mg complexes. The latter term is given by the product of inverse MM constants ($K_{I,q}$) and the concentration of the competing NTP-Mg complexes, $[q]_t$. The parameter $V_{max,i}$ represents the maximum possible elongation rate, while $K_{M,i}$ is the concentration of the NTP-Mg complex that allows for half of the maximum elongation rate to be achieved. 

\item[(4)] \textbf{Termination:} 
In the termination stage, the T7 RNAP enymze reaches to the end of the DNA template, signifying the successful completion of the transcription process. Then, 
the T7 RNAP enzyme dissociates from the DNA template, and both are released at a rate denoted by $\lambda^n(\pmb{s}_t)$, 
    \begin{equation}
    \label{eq:termination_rate}
        \lambda^n(\pmb{s}_t) = V_{max,n} \times \frac{[E\cdot D\cdot M_J]_t}{K_{M,J} + [E\cdot D\cdot M_J]_t}.
    \end{equation}
A full-length RNA transcript with the desired length $J$ is synthesized and released, making it available for downstream applications or analyses. 
This termination rate in (\ref{eq:termination_rate}) is influenced by factors including the concentration of enzyme ($E$), DNA template ($D$), and mRNA ($M_J$), which are all present in the ternary transcription complex. This regulation model contains two parameters: $V_{max,n}$ and $K_{M,J}$, where $V_{max,n}$ represents the maximum possible rate of termination and $K_{M,J}$ represents the concentration of the ternary complex that allows for half of the maximum termination rate to be achieved. Equation (\ref{eq:termination_rate}) is structured to exhibit an increasing termination rate as the concentration of the ternary transcription complex rises. However, there is a saturation point beyond which the termination rate remains at its maximum level.

\item[(5)] \textbf{Degradation:}
During the IVT process, the synthesized full-length RNA product 
inevitably undergoes degradation at a rate $\lambda^d(\pmb{s}_t)$. 
We model the degradation rate $\lambda^d(\pmb{s}_t)$ of RNA at time $t$ as follows:
    \begin{equation}
        \lambda^d(\pmb{s}_t) = (k_{ac}[\mbox{H}]^{n_{ac}}_t + k_{ba}[\mbox{OH}]^{n_{ba}}_t +k_{Mg}[\mbox{Mg}]^{n_{Mg}}_t)[\mbox{RNA}]^{n{RNA}}_t,
        \label{eq:degradation_rate}
    \end{equation}
    where $k_{ac}$, $k_{ba}$, $k_{Mg}$, $n_{ac}$, $n_{ba}$, $n_{Mg}$, and $n_{RNA}$ are model-specific parameters.
The RNA degradation rate in Equation~\eqref{eq:degradation_rate} takes into account the effects from two important environmental factors: the pH and the concentration of magnesium ions (Mg$^{2+}$). The degradation rate is determined by three separate factors: a) the concentration of acidic hydrogen ions (H$^+$) with a corresponding rate constant $k_{ac}$ and exponent $n_{ac}$; b) the concentration of basic hydroxide ions (OH$^-$) with a rate constant $k_{ba}$ and exponent $n_{ba}$; and c) the concentration of magnesium ions with a rate constant $k_{Mg}$ and exponent $n_{Mg}$. The RNA concentration at  time $t$ is also included in the equation with an exponent $n_{RNA}$. 
\end{itemize}

In sum, four types of IVT \textit{reaction rates} are identified and modeled as functions of system state $\pmb{s}_t$ at any time $t$: 1) the binding and initiation rate $\lambda^b(\pmb{s}_t)$; 2) the per-step-elongation rate $\lambda^{e_j}(\pmb{s}_t)$ with $j = 2,3, \dots, J$; 3) the termination rate $\lambda^n(\pmb{s}_t)$; and 4) the degradation rate $\lambda^d(\pmb{s}_t)$. 
The proposed stochastic model for enzymatic molecular reaction networks can capture the inherent stochasticity of the IVT process with respect to reactant concentrations and reaction rates, as well as modeling the complex mechanisms and dynamic interactions of multiple CPPs to improve the prediction of yield and product quality attributes (PQAs). 


\subsection{Macro-kinetic State Transition Modeling}
\label{subsec:stateTransition}

To model the dynamic evolution of the IVT process, at any time $t$, the reaction rates depend on state $\pmb{s}_t$, 
\begin{equation}
    \pmb{\lambda} (\pmb{s}_t) = \{\lambda^b(\pmb{s}_t),\lambda^{e_2}(\pmb{s}_t),\lambda^{e_3}(\pmb{s}_t),\dots,\lambda^{e_J}(\pmb{s}_t), \lambda^n(\pmb{s}_t), \lambda^d(\pmb{s}_t)\}^\top.
    \nonumber
\end{equation}
Let $\pmb{N}$ be a $m \times (J + 2)$ stoichiometry matrix that characterizes the structure of the IVT reaction network with $m$ denoting the dimension of species or state $\pmb{s}_t$. The $(p,q)$-th element of $\pmb{N}$, denoted as $\pmb{N}(p,q)$, represents the number of molecules of the $p$-th species that are either consumed (indicated by a negative value) or produced (indicated by a positive value) in each random occurrence of the $q$-th reaction.

Consequently, during the time interval $[0,t]$, the change in the solution mixture profile is derived 
as:  
\begin{equation}
    {\mathbf{s}}_{t} = {\mathbf{s}}_0 + \pmb{N} \cdot \pmb{R}_t,
    \nonumber
\end{equation}
where $\pmb{R}_t$ is a $(J+2)$-dimensional vector representing the accumulated number of occurrences of each reaction $k$ until time $t$ for $k \in \{b, e_2, e_3, \dots, e_J, n, d\}$, i.e., $\pmb{R}_t = (R^b_t, R^{e_2}_t, R^{e_3}, \dots, R^{e_J}_t, R^n_t, R^d_t)^\top$, and $\pmb{N} \cdot \pmb{R}_t$ represents the net amount of IVT reaction outputs up to time $t$. 
Inspired by \cite{anderson2011continuous}, we assume that the occurrence times $R^k_t$ for each reaction $k$ follows a nonhomogeneous Poisson process. 
The intensity of this process is determined by the reaction rates.
Therefore, the probability that the $k$-th reaction occurs $r$ times during time interval $(t,t+\Delta t]$ becomes,
\begin{equation}\label{eq: nonhomogenuous Poisson}
    P\left(R^k_{t+\Delta t}-R^k_t = r\right) = \frac{e^{-\int_t^{t+\Delta t} \lambda^k (\pmb{s}_x)dx}(\int_t^{t+\Delta t}\lambda^k (\pmb{s}_x)dx)^r}{r!} \triangleq  \text{Poisson}\left(\int_t^{t+\Delta t}\lambda^k (\pmb{s}_x)dx\right).
    \nonumber 
\end{equation}
\vspace{0.1in}

\section{Empirical Study}
\label{sec:empirical}
In this section, simulation experiments were conducted to evaluate the performance of the proposed stochastic molecular reaction queueing network model. The proposed model was first validated in Section~\ref{subsec:Validation} with the data and observations from the study \cite{dousis2023engineered} to ensure that it accurately represents the IVT process. In Section~\ref{subsec:PerformanceAnalysis}, a series of simulation experiments were performed to assess the performance of the IVT system under different conditions.

\subsection{Model Validation}
\label{subsec:Validation}
This section presents the validation of the proposed stochastic mechanistic model in terms of its ability to predict the purity and yield of RNA products by using a batch-based IVT process with a duration of 150 minutes (i.e., $T = 150$). The study conducted by \cite{dousis2023engineered} investigated the IVT process purity profiles by using eight different mRNAs with varying lengths and sequence compositions.  The results revealed that the impurity levels for all eight mRNAs fell within the range of 10\% to 20\%. 
Our model considers two primary sources of impurity: the entry of initiated enzymatic complexes into abortive cycling and the degradation of RNA products.

The kinetic parameters of the proposed model were determined based on the existing research \cite{arnold2001kinetic,van2021quality}. The values of dissociation equilibrium constants were taken to be $10^{-6.95}$ for $K_{eq,1}$ to $K_{eq,4}$, $10^{-4.42}$ for $K_{eq,5}$ to $K_{eq,8}$, $10^{-1.69}$ for $K_{eq,9}$ to $K_{eq,12}$, $10^{-1.49}$ for $K_{eq,13}$ to $K_{eq,16}$, and $10^{-5.42}$, $10^{-2.33}$, $10^{-8.94}$, $10^{-6.13}$, $10^{-3.05}$, $10^{-7.5}$ mol/L for $K_{eq,17}$ to $K_{eq,22}$ respectively. 
The maximum specific reaction rates  $V_{max,\cdot}$ were set as $1.8 \times 10^{-6}$. The Michaelis-Menten affinity constant $K_{M,DNA}$,  $K^b_{M,MgGTP}$, $K_{M,MgATP}$, $K_{M,MgUTP}$, $K_{M,MgCTP}$, $K_{M,MgGTP}$ were determined as $6.3 \times 10^{-9}$, $98 \times 10^{-6}$, $88 \times 10^{-6}$, $44 \times 10^{-6}$, $44 \times 10^{-6}$, and $88 \times 10^{-6}$ respectively. The Michaelis-Menten inhibition constants $K_{I,\cdot}$ were assigned a value of $2 \times 10^{3}$. The parameters related to abortive rate were set as $25 \times 10^{6}$, $0.3 \times 10^{6}$, $0.5$, $0.5$, $0.5$, $0.5$ for $\phi_1$, $\phi_2$, $\phi_3$, $\phi_4$, $\phi_5$, $\phi_6$. For degradation rate, $n_{ac}$, $n_{ba}$, $n_{Mg}$ and $n_{RNA}$ were set to 1 and $k_{ac}$, $k_{bc}$, $k_{Mg}$ were determined as $1.2 \times 10^{6}$, 0, 0 respectively. 

The full length for the target RNA product is set at 50 NTPs. To simplify the analysis, we assume that abortive cycling only occurs when the length of the growing RNA reaches 9 NTPs, considering the fact that the typical length of abortive transcripts ranges from 2 to 10 \cite{dousis2023engineered}. 
The main reason for this simplification is the lack of a comprehensive mechanistic model that accurately characterizes the decrease in abortive rate as the synthesized mRNA length increases.
We conducted $R = 30$ replications of the IVT process and each simulation run has run-length equal to $T  = 150$ minutes. Considering the relatively higher cost of NTPs compared to other raw materials \cite{van2021quality}, we conducted the simulation experiments with limited NTP resources, specifically 0.075~M Mg, 0.0015~M ATP, 0.0015~M UTP, 0.0015~M CTP, 0.0015 M GTP, $10^{-5}$ M DNA template and $10^{-5}$ M T7RNAP. The reactor solution was initially set at a pH level of 8. Figure~\ref{fig:results1} depicts the characteristics of 3 representative samples out of the 30 simulated processes.

\begin{figure}[ht]
	\centering
	\includegraphics[width=\textwidth]{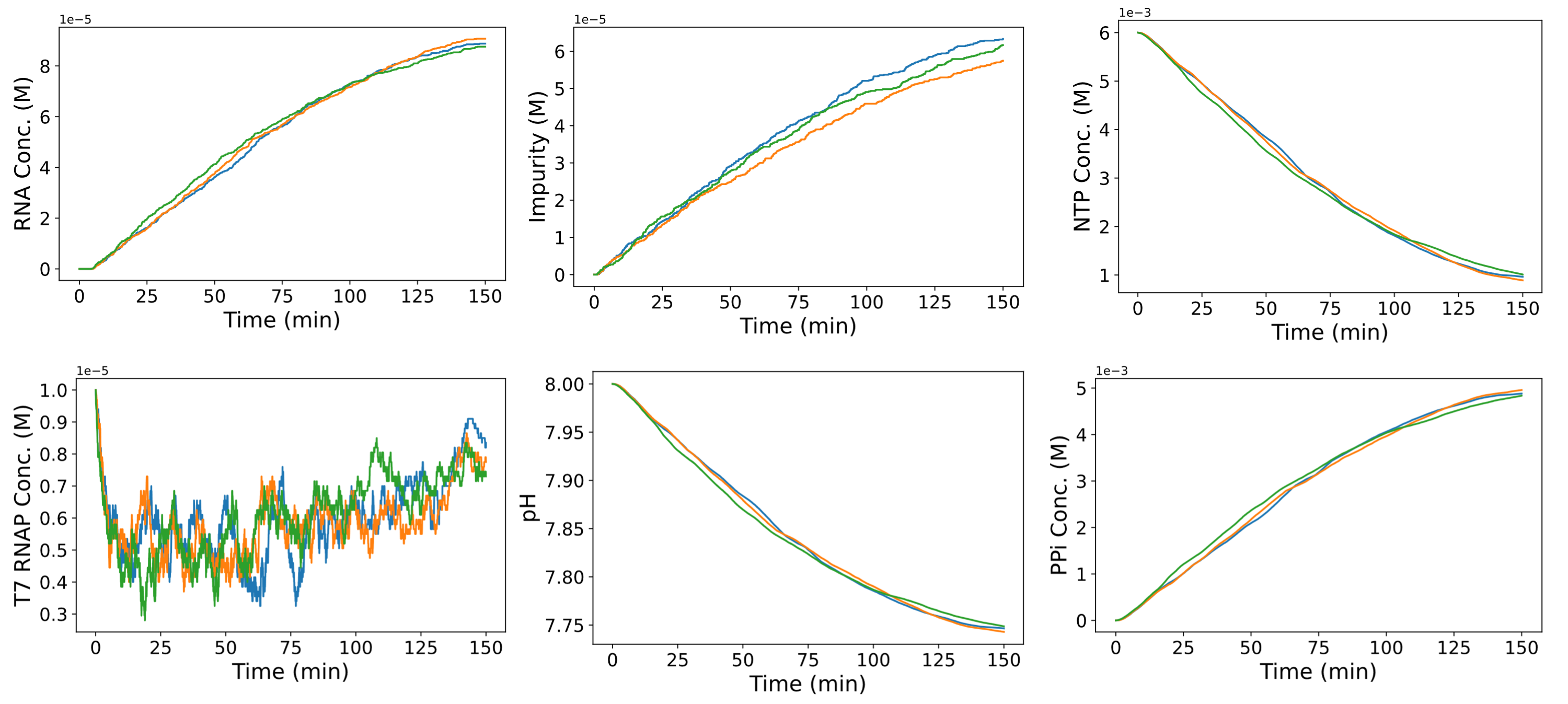}
 \vspace{-0.2in}
	\caption{
	Batch-based IVT process characteristic prediction. (A) Synthesized RNA Concentration, (B) Impurity Concentration, (C) NTP Concentration, (D) T7 RNAP concentration, (E) pH, and (F) PPi concentration. Three colored lines denote three representative simulation output scenarios out of the total 30 replications.
	}
	\label{fig:results1}
\end{figure}

Throughout the process, the production rate of full-length RNA gradually declines due to several reasons. Firstly, the accumulation of PPi leads to a decrease in the availability of the cofactor Mg in the solution. Secondly, the release of hydrogen ions (H) causes T7 RNAP enzyme to deviate from its optimal activity range. Third, the consumption of NTPs, which are raw materials, contributes to the reduction in the rate of RNA synthesis. The expected yield $\eta = \mbox{E}\left[[RNA]_T\right]$ with 95\% confidence interval (CI) across 30 replications was reported as $97.56 \pm 0.82 ~ \mu$M. 
Since the utilized parameters are derived from existing literature estimation results, the reported results maintain validity and reasonableness.
The expected impurity level $\rho = 
\mbox{E} \left[ \frac{9 \times [\mbox{Impurity}]_T}{9 \times [\mbox{Impurity}]_T + 50 \times [\mbox{RNA}]_T}
\right]$ with 95\% CI was reported as $11.23 \pm 0.19 \%$ weight/weight (w/w), which is consistent with the findings (i.e., 10\% to 20\%) in \cite{dousis2023engineered}. 

\subsection{IVT Process Performance Analysis under Different Decisions}
\label{subsec:PerformanceAnalysis}

In this section, we examine the performance of the IVT process under different decision scenarios for the CPPs, considering the CPP-CQA relationship summarized in Table~\ref{table:CPP_and_CQA}. It is worth noting that T7 RNAP, Mg, and NTPs play vital roles as raw materials.  Insufficient concentrations of these components can result in inefficient IVT process performance. Conversely, excessively high concentrations of NTPs or imbalanced proportions among different types of NTPs can result in substrate nucleotide competition. This leads to a decrease in the transcription reaction rate. Therefore, we focus on the pH in the transcription reactor and the concentration of NTPs. We conducted 30 replications ($R = 30$) for each experimental setting.

\begin{itemize}
    \item [(1)] \textbf{pH:} The model was implemented at different pH levels, specifically 7, 7.5, 8, 8.5, and 9. All other experimental parameters remained the same as described in Section~\ref{subsec:Validation}. The performance of the IVT process, with respect to the RNA yield and impurity level, is presented in Table~\ref{tab:pH_result}.
    
    \begin{table}[h!]
    \centering
    \caption{Batch-based IVT production system performance at different pH conditions.}
    \label{tab:pH_result}
    \begin{tabular}{|c|ccccc|}
    \hline
    pH                      & 
    7             & 7.5           & 8             & 8.5           & 9 \\ \hline
    Yield ($\mu$M) & 
    $24.38 \pm 0.74$ & $84.75 \pm 0.38$ & $97.56 \pm 0.82$ & $95.44 \pm 0.62$ & $71.63 \pm 0.50$ \\ 
    Impurity  ($\%$)         & 
    $56.01 \pm 0.90$ & $16.48 \pm 0.22$ & $11.23 \pm 0.19$ & $12.72 \pm 0.03$ & $23.27 \pm 0.18$   \\ \hline
    \end{tabular}
    \end{table}
    
    The efficiency of T7 RNAP is significantly influenced by the reactor environment, and the optimal condition is pH$\approx 8$. Under this condition, the process achieves the highest yield of $97.56 \pm 0.82$ with the lowest impurity level of $11.23 \pm 0.19$. When the pH is maintained within the range of 7.5 to 8.5, the system demonstrates comparable performance. However, when the pH deviates from this range, such as pH equal to 7 or 9, the activity of the enzyme declines rapidly. As a result, the production rate decreases significantly, and impurities accumulate due to the elevated abortive rate.
    
    \item [(2)] \textbf{NTPs:} The simulation experiments were performed using different concentrations of NTPs, namely 0.006~M, 0.008~M, 0.01~M, and 0.012 M, with each NTP set at an equal concentration. The results provide insights into the system's performance, specifically in terms of RNA yield and impurity level, as summarized in Table~\ref{tab:NTP_result}.

    \begin{table}[h!]
    \centering
    \caption{Batch-based IVT production system performance at different NTP concentrations.}
    \label{tab:NTP_result}
    \begin{tabular}{|c|cccc|}
    \hline
    NTP (M)         & 0.006 & 0.008 & 0.01 & 0.012 \\ \hline
    Yield ($\mu$M) & $97.56 \pm 0.82$ & $118.88 \pm 0.77$ & $131.38 \pm 1.96$ & $134.38 \pm 1.46$  \\ 
    Impurity  ($\%$) & $11.23 \pm 0.19$ & $10.95 \pm 0.13$ & $10.56 \pm 0.38$ & $10.75 \pm 0.20$  \\ \hline
    \end{tabular}
    \end{table}

    The yield exhibits an increase (from $97.56 \pm 0.82$ to $134.38 \pm 1.46$) as the NTPs concentrations increase, emphasizing their crucial role in the IVT process. However, the rate of increase diminishes due to two potential reasons: a) the higher NTPs concentrations lead to increased competition inhibition \cite{arnold2001kinetic}, which is incorporated into the regulation mechanistic model in Equations~(\ref{eq:binding_rate}) and (\ref{eq:elongation_rate}); and b) the concentrations of Mg, DNA template, or T7 RNAP become limiting factors, hindering further improvements. Conversely,  the impurity level remains relatively stable and does not show significant changes across the range of NTPs concentrations.
\end{itemize}

\section{Conclusion}
\label{sec: conclusion}
The urgency of rapid vaccine production has been highlighted by recent viral outbreaks. RNA vaccines, known for their advantages in terms of speed, effectiveness, and safety, offer a promising solution. 
In order to tackle the manufacturing challenges associated with RNA vaccine production, this study develops a mechanistic model that can support the prediction and analysis for the IVT process. This model combines a novel stochastic molecular reaction queueing network with a regulatory kinetic model characterizing the influence of system state (e.g., pH, the concentrations of DNA template, NTPs, magnesium, and T7 RNAP) on reaction rates and production outputs such as RNA yield 
and impurity levels.
The empirical study shows that this model has a promising performance across different process conditions. Ultimately, it demonstrates the potential to enhance RNA product quality consistency, increase yield, and lower production costs. 



\footnotesize

\bibliographystyle{unsrt}

\bibliography{proposal,proj_ref, eduardo_extra}

\end{document}